\documentclass[twocolumn,twoside,slac_two]{revtex4}
\usepackage{graphicx}
\usepackage{fancyhdr}
\begin{document}

\title{Unraveling the Emission Geometry of the \textit{Fermi} Millisecond Pulsars}

%

\author{C. Venter,$^{1,2,3}$ A.K. Harding,$^1$ and L. Guillemot$^4$}
\affiliation{$^1$Astrophysics Science Division, NASA Goddard Space Flight Center, Greenbelt, MD 20771, USA}
\affiliation{$^2$Unit for Space Physics, North-West University, Potchefstroom Campus, Private Bag X6001, Potchefstroom 2520, South Africa}
\affiliation{$^3$NASA Postdoctoral Program Fellow,}
\affiliation{$^4$Max-Planck-Institut f\"{u}r Radioastronomie, Auf dem H\"{u}gel 69, 53121 Bonn, Germany}


\begin{abstract}
The nine millisecond pulsars (MSPs) that have now been detected by \textit{Fermi}-LAT are providing an excellent opportunity to probe the emission geometry of these ancient compact objects. As they are radio-loud, one may use the relative phase lags across wavebands to obtain constraints on the orientation, size, and location of their radio and gamma-ray beams. We model the gamma-ray light curves using geometric outer gap (OG) and two-pole caustic (TPC) models, in addition to a pair-starved polar cap (PSPC) model which incorporates the full General Relativistic E-field. We find that most MSP light curves are fit by OG and TPC models, while PSPC is more appropriate for two others. The light curves of the newest discovery, PSR~J0034$-$0534, are best modeled using outer magnetosphere OG / TPC models of limited extension for both radio and gamma-ray beams. We model the radio emission of the other eight MSPs using a fixed-altitude conal model at lower altitude. We lastly deduce values for inclination and observer angles ($\alpha$ and $\zeta$), as well as the flux correction factor, in each case.
\end{abstract}

\maketitle

\thispagestyle{fancy}

\section{INTRODUCTION\label{Intro}}
Our pursuit of understanding gamma-ray pulsars has been dramatically furthered by the detection of 46~pulsars after only six months of observation by the \textit{Fermi} Large Area Telescope (LAT)~\cite{Abdo09_Cat}. Owing to the \textit{Fermi} LAT's superior capabilities compared to its predecessor, the \textit{Energetic Gamma Ray Experiment Telescope (EGRET)}, a subpopulation of gamma-ray millisecond pulsars (MSPs) is now emerging~\cite{Abdo09_MSP}, twenty seven years after the discovery of the first MSP~\cite{Backer82}. This population includes PSR~J0218+4232 (first marginally detected by \textit{EGRET}~\cite{Kuiper00}), the first \textit{Fermi}-detected MSP PSR~J0030+0451, and now a ninth MSP PSR~J0034$-$0534 which has nearly phase-aligned radio and gamma-ray light curves~\cite{Abdo09_J0034}. It is expected that this number will soon grow as \textit{Fermi} accumulates data. Additionally, follow-up radio observations of unidentified \textit{Fermi} sources are also uncovering new MSPs~\cite{Ransom09}.

The availability of quality gamma-ray and radio MSP data now affords the opportunity to investigate the mechanisms and locations of emission in MSP magnetospheres. Several different approaches have been followed to describe the high-energy (HE) radiation. Polar cap (PC) models~\cite{DH96} assume near-surface emission along with magnetic pair production. This establishes a low-altitude pair formation front which screens the accelerating electric field~\cite{HM98}. Slot gap (SG) models~\cite{Arons83,MH03_SG} envision the formation of narrow acceleration gaps in a two-pole caustic (TPC) geometry~\cite{Dyks03}, extending from the neutron star (NS) surface to near the light cylinder, along the last open B-field lines. Lastly, outer gap (OG) models~\cite{Romani96} assume that photon-photon pair production-induced cascades produce HE emission along the last open field lines above the null-charge surfaces (where the Goldreich-Julian charge density changes sign). Modern extensions to this model include a 3D solution~\cite{Hirotani08} where a small positive acceleration E-field reaches to the NS surface.

We model these ancient, recycled MSPs~\cite{Bhattacharya91} using 3D emission modeling, including Special Relativistic effects of aberration and time-of-flight delays, and rotational sweepback of B-field lines~\cite{Venter_MSP09}. In addition to using geometric OG and TPC models, we find that some light curves exclusively favor pair-starved polar cap (PSPC) pulsar models, where the pair multiplicity is not high enough to screen the accelerating E-field and charges are continually accelerated up to high altitudes over the full open-field-line region~\cite{HUM05,Frackowiak05a,VdeJ05}. For this case, we implement a solution of the acceleration E-field valid up to high altitudes~\cite{MH04_PS}. Our calculation of light curves and flux correction factors for the case of MSPs is therefore complementary to another work~\cite{Watters09} which focuses on younger pulsars, although our OG and TPC models have non-zero emission widths. We describe the details of the various models in Section~\ref{sec:Model}, and present our results and conclusions in Sections~\ref{sec:Results} and~\ref{sec:Con}.

\section{MODEL DESCRIPTION\label{sec:Model}}
Here we briefly describe the models used to generate the radio and gamma-ray light curves. A detailed description may be found elsewhere~\cite{Venter_MSP09}. In addition, PSR~J0034$-$0534 presents a special case, and its modeling will be fully discussed in another publication~\cite{Venter10}.

\subsection{Magnetic Field}
We use an implementation~\cite{Dyks04_B} of the retarded vacuum dipole B-field solution~\cite{Deutsch55}, assuming that it reasonably approximates the MSP magnetospheric structure. This B-field geometry is characterized by an asymmetrical PC shape due to rotational sweepback of the field lines. We assume constant emissivity~\cite{Dyks03} along the B-lines in the gap regions of the geometric OG and TPC models, while we calculate this explicitly in the PSPC model using the General Relativistic E-field (Section~\ref{sec:E}). 

To calculate the direction of photon propagation~\cite{Romani95}, we find the tangent to the local B-line in the co-rotating frame (using the fact that the vacuum B-field is approximately the same in this frame as in the inertial observer frame), and then aberrate this direction via a Lorentz transformation (transforming from the instantaneously co-moving frame to the observer frame). We also include phase shifts due to rotational sweepback of the B-field lines and travel time delays due to the finite speed of light. Our calculations are consistent to first order in $r/R_{LC}$, with $r$ the radial position, and $R_{\rm LC}=c/\Omega$ the light cylinder radius, with $\Omega$ the rotational velocity of the MSP. Future work may include higher-order corrections, especially when considering the force-free magnetospheric structure~\cite{BS09}. We furthermore explicitly use the curvature radius of the B-field lines as calculated in the inertial observer frame when calculating curvature radiation (CR) losses for the PSPC case. We calculated light curves from OG and TPC models for several different gap widths $w$, while we used $w=1$ for the PSPC model (Section~\ref{sec:Results}).

\subsection{Accelerating Electric Field\label{sec:E}}
In the case of the PSPC model, we only consider CR losses suffered by electron primaries as they move along the B-field lines when modeling the HE emission. Previous studies~\cite{VdeJ05,Frackowiak05a,HUM05} have used a solution~\cite{HM98} of the E-field parallel to the B-field ($E_{||}$), which is valid up to a significant fraction of $R_{\rm LC}$, for the PSPC E-field. We now incorporate the small-angle-approximation solution of $E_{||}$ valid for altitudes very close to $R_{\rm LC}$~\cite{MH04_PS}. Our implementation of $E_{||}$ conserves energy along the relativistic electron primaries' trajectories, with electric potential energy being converted into gamma-ray radiation and particle kinetic energy.

\subsection{Radio Emission Model}
For eight of the \textit{Fermi} MSPs (excluding PSR~J0034$-$0534), we model the radio emission beam using an empirical cone model which assumes that the radio emission originates in a cone of fixed altitude~\cite{Kijak03}, centered on the magnetic axis. We adopt a description~\cite{Harding07_Geminga} based on fits~\cite{ACC02} of average-pulse profiles of a small collection of pulsars at 400~MHz to a core and single cone beam model. The emission occurs increasingly close to $R_{\rm LC}$ as the period $P$ decreases (for more or less constant $\dot{P}$), and is typically located at $10\% - 20\%$ of $R_{\rm LC}$ for these MSPs. Depending on where the observer's line-of-sight intersects the radio cone, the radio profile may exhibit zero, one or two peaks.

In contrast, the near phase alignment of the radio and gamma-ray light curves of PSR~J0034$-$0534 forces us to depart from the conal framework when modeling this newest MSP member. The data seem to require caustic origin of the radio (see Section~\ref{sec:J0034}).
\renewcommand{\thefootnote}{\alph{footnote}}
\begin{table*}[t]
\begin{center}
\caption{Inferred values of $\alpha$, $\zeta$, and $f_\Omega(\alpha,\zeta,P)$}
\begin{tabular}{lcccccccccccc}
\hline 
PSR Name & \textbf{$\alpha_{\rm TPC}$} & \textbf{$\zeta_{\rm TPC}$} & \textbf{$\alpha_{\rm OG}$} & \textbf{$\zeta_{\rm OG}$} & \textbf{$\alpha_{\rm PSPC}$} & \textbf{$\zeta_{\rm PSPC}$} & \textbf{$\alpha_{\rm radio}$} & \textbf{$\zeta_{\rm radio}$} & Ref. & \textbf{$f_{\rm \Omega,TPC}$} & \textbf{$f_{\rm \Omega,OG}$} & \textbf{$f_{\rm \Omega,PSPC}$}\\
\hline 
J0030+0451   & 70$^\circ$      & 80$^\circ$      & 80$^\circ$      & 70$^\circ$      & --- & --- & $\sim62^\circ$      & $\sim72^\circ$          &~\cite{Lommen00}   & 1.04    & 0.90    & ---\\
J0034$-$0534       & 30$^\circ$\tablenotemark[1]      & 70$^\circ$\tablenotemark[1]      & 30$^\circ$\tablenotemark[2]      & 70$^\circ$\tablenotemark[2]  & --- & --- & ---      & ---                &                 & 0.74\tablenotemark[1]    & 0.45\tablenotemark[2]    & ---\\
J0218+4232   & 60$^\circ$      & 60$^\circ$      & 50$^\circ$      & 70$^\circ$      & --- & --- & $\sim8^\circ$       & $\sim90^\circ$          &~\cite{Stairs99}   & 1.06    & 0.63    & ---\\
J0437$-$4715 & 30$^\circ$      & 60$^\circ$      & 30$^\circ$      & 60$^\circ$      & --- & --- & $20^\circ-35^\circ$       & $16^\circ-20^\circ$           &~\cite{MJ95,Gil97} & 1.23    & 1.82    & ---\\
J0613$-$0200 & 30$^\circ$      & 60$^\circ$      & 30$^\circ$      & 60$^\circ$      & --- & --- & small $\beta$ & ---           &~\cite{Xilouris98}   & 1.19    & 1.76    & ---\\
J0751+1807   & 50$^\circ$      & 50$^\circ$      & 50$^\circ$      & 50$^\circ$      & --- & --- & ---       & ---           &     & 0.80    & 0.65    & ---\\
J1614$-$2230 & 40$^\circ$      & 80$^\circ$      & 40$^\circ$      & 80$^\circ$      & --- & --- & ---       & ---           &     & 0.83    & 0.64    & ---\\
J1744$-$1134 & --- & --- & --- & --- & 50$^\circ$  & 80$^\circ$      & --- & --- &     & --- & --- & 1.19\\
J2124$-$3358 & --- & --- & --- & --- & 40$^\circ$  & 80$^\circ$      & $20^\circ-60^\circ$ (48$^\circ$)  & $27^\circ-80^\circ$ (67$^\circ$)      &~\cite{Manchester04}   & --- & --- & 1.29\\
\hline
\end{tabular}
\label{tab2}
\end{center}
\footnotetext[1]{Limited TPC model.}
\footnotetext[2]{Limited OG model.}
\end{table*}
\renewcommand{\thefootnote}{\arabic{footnote}}

\subsection{Flux Correction Factor}
\label{sec:fom}
In order to convert the observed (phase-averaged) energy flux $G_{\rm obs}$ to the all-sky luminosity $L_\gamma$, which defines the gamma-ray conversion efficiency $\eta_\gamma\equiv L_\gamma/\dot{E}_{\rm rot}$, with $\dot{E}_{\rm rot}$ the spin-down luminosity, we need to calculate a flux correction factor ($f_\Omega$). This factor accounts for the fact that an observer only sees a small part of the total radiation: that coming from a slice through the emission beam. The flux correction factor is defined as follows~\cite{Watters09}:
\begin{equation}
 L_\gamma = 4\pi f_\Omega d^2 G_{\rm obs},
\end{equation}
\begin{equation}
f_\Omega(\alpha,\zeta_E) = \frac{\int\!\!\!\int F_\gamma(\alpha,\zeta,\phi)\sin\zeta d\zeta d\phi}{2\int F_\gamma(\alpha,\zeta_E\phi)\,d\phi},\label{eq:fom}
\end{equation}
with $F_\gamma$ the photon flux per solid angle (`intensity'), and $\zeta_E$ the Earth line-of-sight (assuming similar distributions of gamma-ray photon and energy fluxes in $(\zeta,\phi)$-space, i.e., $F_{\gamma}(\zeta,\phi)/F_{\rm \gamma,tot}\approx G_{\gamma}(\zeta,\phi)/G_{\rm \gamma,tot}$, with $\phi$ the phase). Results of calculations of $f_\Omega$ for different pulsar models appear in Table~\ref{tab2}.

\section{RESULTS\label{sec:Results}}
\begin{figure*}[t]
\centering
\includegraphics[width=136mm]{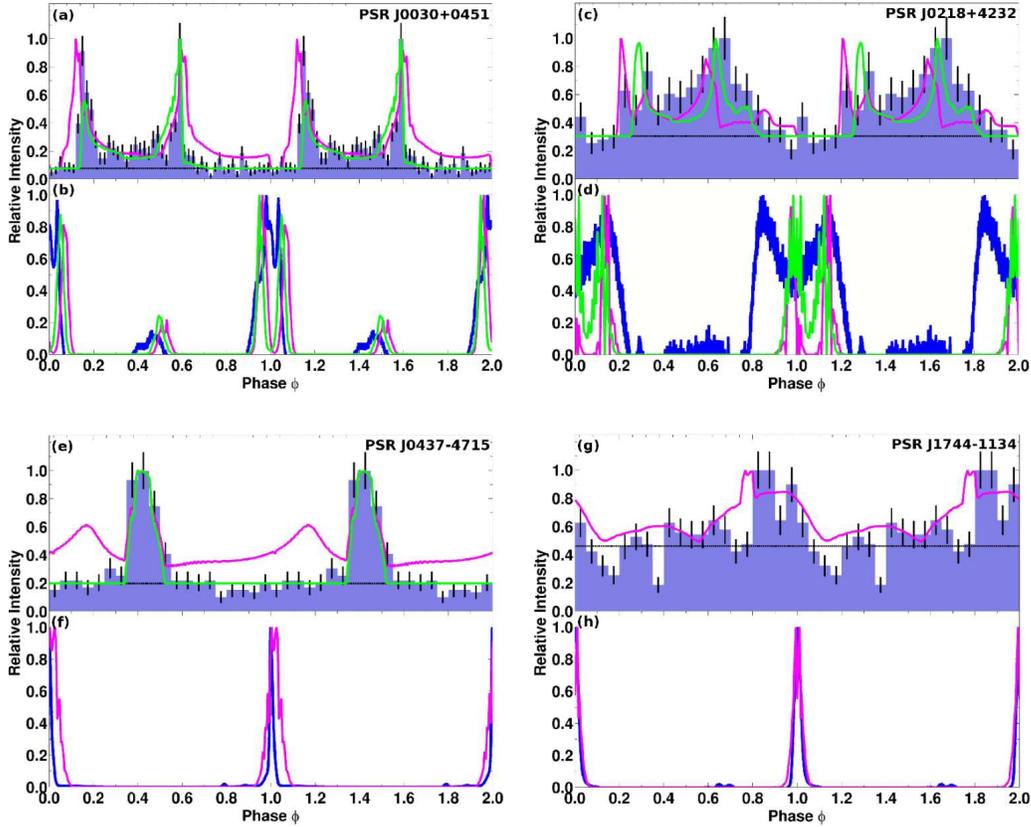}
\caption{Plots of the observed and fitted gamma-ray and radio light curves for several MSPs: PSR~J0030+0451, PSR J0218+4232, PSR J0437$-$4715, and PSR~J1744$-$1134. In each case, in the upper panels the histograms represent the \textit{Fermi}-LAT data above 0.1~GeV~\citep{Guillemot09}, the horizontal dashed line the estimated background level, the green lines are OG fits, and magenta lines are TPC fits (see Table~\ref{tab2}). The bottom panels indicate radio light curves, where blue lines represent the radio data, while the magenta and green lines correspond radio model fits with the same $(\alpha,\zeta)$ combinations as those of the respective OG and TPC fits. (For the last two radio panels, the OG and TPC models are for the same $\alpha$ and $\zeta$, giving a unique radio conal beam fit.) Light curves were chosen to show different classes of gamma-ray models: panel~(a) and~(c) indicate OG / TPC models with large $\alpha$, panel~(e) shows OG / TPC models with small $\alpha$, and panel~(g) shows a PSPC model.} \label{fig1}
\end{figure*}

We generated light curves for each of the different pulsar models, using a range of inclination and observer angles ($\alpha = \zeta = 5^\circ-90^\circ$, in $5^\circ$ intervals), periods $P=2$, 3, and~5~ms, and gap widths $w$ (ranging from $0.05-0.4$ for the OG / TPC models, and $w=1$ for the PSPC case). We matched model light curves to the MSP gamma-ray and radio data by eye (although statistical uncertainties in the data as well as similar predicted light curves shapes for similar ranges of $\alpha$ and $\zeta$ may complicate unique matching). Figure~\ref{fig1} shows an example of this for four MSPs. We did not find any satisfactory fits for the geometric PC models or for OG and TPC models with $w=0$. In addition, the radio cone model fits the data (excluding PSR~J0034$-$0534) quite well overall, except for the case of PSR~J0218+4232, which seem to require a wider cone beam. The MSP radio beam may be relatively large, as its total size and annular width scales as $P^{-0.35}$ in our model.

From our light curves fits, we infer values for $\alpha$ and $\zeta$ for each MSP. These are summarized in Table~\ref{tab2} (labeled with subscripts `TPC', `OG', and `PSPC'), and compared with values obtained from radio polarimetric measurements (labeled with subscripts `radio'). Our calculated values for $f_\Omega(\alpha,\zeta,P)$ indicate that this factor is typically of order unity for the best-fit geometries we consider here, with the OG model usually predicting lower values than the TPC model. Our results qualitatively resemble other calculations for OG / TPC models applied to young pulsars~\cite{Watters09}, while the difference encountered when using the unscreened PSPC model (vs.\ the PC model) reflects the fundamental physical dissimilarity of the magnetospheric structure of pair-starved MSPs and younger pulsars.

\subsection{OG / TPC Case}
Both OG and TPC models have a preponderance of double-peaked light curves at similar phases. Sharp, solitary peaks exist for some regions in phase space in OG models, while the corresponding TPC-peaks usually have additional low-level features. No emission originates below the null charge surface ($\zeta=90^\circ$) in the OG model, implying that emission is only visible from one magnetic pole, in contrast to the TPC model where emission is visible from both poles. Consequently, OG models do not exist at all angle combinations, while TPC models do. (We impose a maximum emission radius $r_{\rm max}=1.2R_{\rm LC}$ for these models, together with a maximum cylindrical radius $\rho_{\rm max}=0.95R_{\rm LC}$, while we assume that the emission starts at the stellar surface $r=R$.)

Six MSPs (all in Table~\ref{tab2} excluding PSR~J0034$-$0534, PSR~J1744$-$1134, and PSR J2124$-$3358) are well fit by OG and TPC models. For these cases, the gamma-ray light curve lags the radio. In addition, PSR~J0030+0451, PSR~J0218+4232, and PSR~J1614$-$2230 (as well as PSR~J0034$-$0534; see Section~\ref{sec:J0034}) exhibit double-peaked light curves, which signifies the existence of screening electron-positron pairs that form the OG or TPC emitting structure.

\subsection{PSPC Case}
For the PSPC model, we find mostly single-peaked gamma-ray profiles roughly in phase with the radio (for single radio peaks), which increase in width when $P$ decreases, especially for the radio. For large impact angles $\beta = \zeta-\alpha$, only gamma-ray radiation is visible, since the gamma-ray beams are believed to be larger than the radio ones (possibly explaining the phenomenon of `radio-quiet' pulsars~\cite{Abdo09_Cat}). 

Double-peaked profiles occur only for large $\zeta$ (not for both large $\alpha$ and $\zeta$ as in a constant-emissivity PC model), because the presence of `favorably-curved B-field lines' at large~$\alpha$ is reflected in the azimuthal dependence of $E_{||}$. Emission is therefore usually only visible from a single magnetic pole, as in the OG case. The gamma-ray emission originates on all open field lines, while the radio originates at a fixed height. The gamma-ray profile therefore appears at earlier phases than the radio (with the caustic peaks being washed out), so that the radio peak lags the gamma-ray peak. This is the case for PSR~J1744$-$1134 and PSR~J2124$-$3358, which are exclusively fit by the PSPC model. For these MSPs, the gamma and radio emission originate from the same magnetic pole, well above the NS surface.

\subsection{Limited OG / TPC Case\label{sec:J0034}}
\begin{figure*}[t]
\centering
\includegraphics[width=95mm]{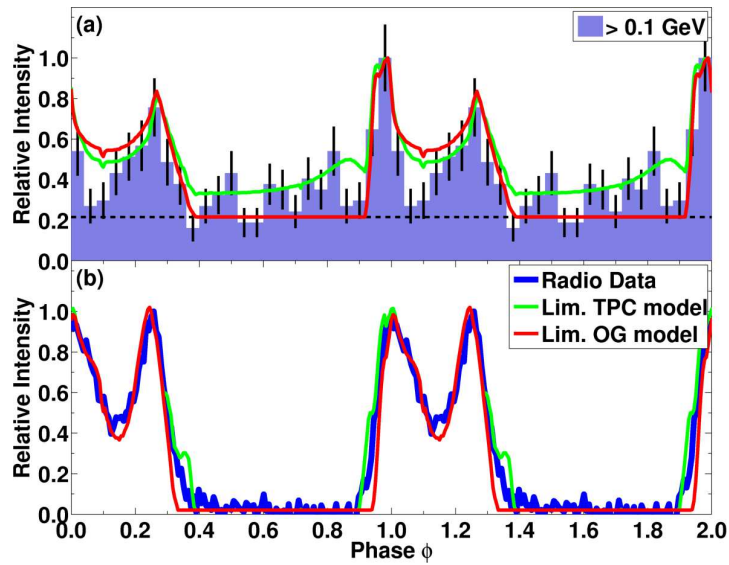}
\caption{Data and model light curves for PSR~J0034$-$0534. The top histogram indicates \textit{Fermi}-LAT data above 0.1~GeV~\cite{Abdo09_J0034} along with limited TPC (green) and limited OG (red) fits. Similar for the bottom panel, where the solid blue line is the Westerbork Synthesis Radio Telescope (WSRT) 324~MHz radio profile. We used $\alpha = 30^\circ$, $\zeta = 70^\circ$, and $w = 0.05$ for all light curves.}\label{fig2}
\end{figure*}

We find that a combination of a radio PC model and a gamma-ray OG or TPC model does not reproduce PSR~J0034$-$0534's light curves~\cite{Venter10}. Rather, a co-located caustic origin of both radio and gamma-ray emission is probably required~\cite{Abdo09_J0034}. The so-called `limited OG / TPC' models (Figure~\ref{fig2}) refer to OG and TPC models where we have limited the extent of the emission regions along the B-field lines, with the radio subsumed into the more extended OG / TPC regions. 

The minimum and maximum emission radii are quite well constrained by the profile shapes and radio-to-gamma phase lags. We found good fits for $\alpha=30^\circ$, $\zeta=70^\circ$, $w=0.05$, and using $(r_{\rm min},r_{\rm max})=(0.12,0.9)R_{\rm LC}$ for the gamma-ray profile, while assuming $(r_{\rm min},r_{\rm max})=(0.6,0.8)R_{\rm LC}$ for the radio profile. This is the first time that near phase alignment of the radio and gamma-ray pulses is seen for an MSP, similar to what is observed in the Crab pulsar.

\section{DISCUSSION AND CONCLUSIONS\label{sec:Con}}
The gamma-ray observations now support a stronger `familial' link between the MSP and young pulsar classes in the sense that we observe spectral and temporal characteristics in their HE emission which are very much alike. This indicates the operation of very similar emission mechanisms regardless of the aeons separating the different pulsar generations. Interestingly, it is noted~\cite{Abdo09_Cat} that the MSPs and young pulsars observed by \textit{Fermi} share magnetic fields at the light cylinder which are comparable in magnitude, possibly providing further corroboration for this idea (see also~\cite{Cognard96}). Also, explanation of the surprising inference that there must be a significant number of pairs in MSP magnetospheres to support the OG / TPC framework, may require a B-field structure departing significantly from the usual dipole picture~\cite{Chen93,Lamb09}. Thus, the surface B-fields of younger pulsars and MSPs may be of similar magnitude (unless the younger pulsars' B-fields are also boosted, but by a different mechanism than in the MSP case), and the low-altitude B-field structure should play an important role in the magnetic pair creation physics encountered in TPC and PC models. Additionally, recent measurements indicate large MSP masses of up to $\sim1.7M_\odot$~\cite{Freire09}, larger than the average mass of young pulsars, and this may impact MSP B-field properties and enhance pair creation probability.

We were able to demonstrate a bifurcation in the HE MSP population, distinguishing between MSPs with light curves modeled by the OG / TPC models on the one hand, and those fit by the PSPC model on the other hand. This is due to the fact that we modeled both the gamma-ray and radio curves for each pulsar. Both the shapes and the relative radio-to-gamma phase lags provided by the data could then be used to constrain the best-fit model type and geometry for each MSP. 

However, PSR~J0034$-$0534 respresents a third MSP subclass, as we inferred that an outer magnetosphere caustic origin of both the gamma-ray and radio emission is required to explain its nearly phase-aligned light curves. We therefore conclude that the HE emission must come from the outer magnetosphere in \textit{all} MSP models considered, as is also believed to be true for the bulk of the gamma-ray pulsar population~\cite{Abdo09_Cat}. 

Our calculations of the flux correction factor imply a wide beaming angle, which derives from the fact that we obtain best fits for large impact angles. The larger radio beam widths of MSPs compared to those of canonical pulsars furthermore suggests that there should be relatively few radio-quiet MSPs.

Studies involving phase-resolved spectroscopy of \textit{Fermi}-LAT data, as well as the anticipated discovery of several more gamma-ray pulsars, should make for a vibrant field of research for the foreseeable future.

\bigskip 
\begin{acknowledgments}
C.V.\ is supported by the NASA Postdoctoral Program at the Goddard Space Flight Center, administered by Oak Ridge Associated Universities through a contract with NASA, and also by the South African National Research Foundation. A.K.H.\ acknowledges support from the NASA Astrophysics Theory Program. 
\end{acknowledgments}

\bigskip 

\end{document}